\documentclass[11pt,fleqn]{article}

\usepackage{amsmath,amssymb}
\usepackage{graphicx}
\usepackage[top=0.75in, bottom=0.75in, left=0.65in, right=0.65in, dvips]{geometry}

\begin{document}

\title{{\sf Trailing Edge Unification via an Intermediate Pati-Salam Group}}

\author{V.\ Elias\thanks{Department of Applied Mathematics, The University of Western Ontario, London, Ontario  N6A 5B7, Canada} ~~and
T.G.\ Steele\thanks{Department of Physics and Engineering Physics, University of
Saskatchewan, Saskatoon, SK, S7N 5E2, Canada}
}

\maketitle

\begin{abstract}We demonstrate to two-loop order that an intermediate symmetrically embedded Pati-Salam $SU(2)_L \times SU(2)_R \times SU(4)$ level of symmetry is all that is necessary to accommodate empirical values of $\alpha\left(M_z\right), \alpha_s\left(M_z\right)$ and $\sin^2\theta_w \left(M_z\right)$ within a grand unification context but with a high ($10^{14}\,{\rm GeV}$) intermediate mass scale and with a concomitant higher GUT scale.
\end{abstract}

The Pati-Salam Model \cite{1} served to explain the generational structure of quarks and leptons by proposing that lepton number be the fourth colour.  Although it is generally known for its proposal of integer charged quarks, it could certainly be formulated with conventional charges for the quarks and leptons, and  the phenomenological case for Pati-Salam has been made elsewhere \cite{moha}. Back in the time when $\sin^2 \theta_w$ was thought to be much larger than its empirical value today, it was possible to show how an intermediate level of Pati-Salam symmetry could allow values of $\sin^2 \theta_w$ as high as 3/8 at low energies \cite{2}.

In the present work, we demonstrate how present empirical values of $\alpha_s\left(M_z\right)$, $\alpha\left(M_z\right)$ and $\sin^2 \theta_w \left(M_z\right)$ can be fully accommodated by an intermediate level of Pati-Salam Symmetry within a Grand Unified Context.
We begin by considering the neutral current sector of the Pati-Salam group $SU(2)_L \times SU(2)_R \times SU(4)_c$.  There are four chiral $SU(2)_L$ and $SU(2)_R$ doublets $\Psi_L^a$ and $\Psi_R^a$ per generation:  {\it i.e.,} for $SU(2)_L$ we have $\Psi_{LR}^a = (u_R, d_R)_L$, $\Psi_{L_Y}^a = (u_Y, d_Y)_L$, $\Psi_{LB}^a = (u_B, d_B)_L$, $\Psi_{L\ell}^a = (v_e, e^-)_L$. Similarly there are two non-chiral fundamental representations $\Phi^a$ per generation of $SU(4)_c$:  $\Phi_1^a = (u_R, u_Y, u_B, v_e)$ and $\Phi_2^a = (d_R, d_Y, d_B, e^-)$.  The neutral current sector (discounting the eight $SU(3)$ gluons $V_1 \cdots V_8$) is obtained from the Lagrangian current
\begin{equation}
\begin{split}
\Delta {\cal{L}} & =  \sum_{K=1}^4 \overline{\Psi}_{LK}^a i\gamma^\mu \left[ \partial_\mu + \frac{ig_L}{2} \left( 
\begin{array}{clcr}
W^3_{L_\mu} & \\
 & -W^3_{L_\mu}
 \end{array}
 \right) \right]_{ab} \Psi_{LK}^b
 \\
& +  \sum_{K=1}^4 \overline{\Psi}_{RK}^a i\gamma^\mu \left[ \partial_\mu + \frac{ig_R}{2} \left( 
\begin{array}{clcr}
W^3_{R_\mu} & \\
& -W^3_{R_\mu}
\end{array}
\right) \right]_{ab} \Psi_{RK}^b  
\\
& + \sum_{I=1}^2 \overline{\Phi}_I^a i\gamma^\mu \left[ \partial_\mu + \frac{ig_4}{\sqrt{24}} \left( 
\begin{array}{clcr}
1 & & & \\
& 1 & & \\
& & 1 & \\
& & & -3
\end{array}
\right) V_\mu^{15} \right]_{ab} \Phi_I^b\quad . 
\end{split}
\label{eq1}
\end{equation}
For fractional quark charges one finds that the photon $A_\mu$ and the normal Weinberg-Salam $Z_\mu$ mediating $SU(2)_L \times U(1)$ neutral currents are given by
\begin{gather}
A^\mu = W_3^{L\mu} \sin\theta_w + W_3^{R\mu} \cos\theta_w \cos\phi + V_{15}^\mu \cos\theta_w \sin\phi
\\
Z^\mu = W_3^{L\mu} \cos\theta_w - W_3^{R\mu} \sin\theta_w \cos\phi - V_{15}^\mu \sin\theta_w \sin\phi
\end{gather}
and that the remaining neutral gauge boson not present in the $SU(2) \times U(1)$ subgroup is
\begin{equation}
Z^{\prime\mu} = W_3^{R\mu} \sin\theta_w \sin\phi - V_{15}^\mu \sin\theta_w \cos\phi ~.
\end{equation}
Correct charge assignments for fermions coupling to the photon are obtained by requiring that
\begin{equation}
g_L \sin\theta_w = e, \; g_R \cos\theta_w \cos\phi = e,\;
g_4 \cos\theta_w \sin\phi = \sqrt{2/3} \, e ~.
\label{eq2}
\end{equation}
One then sees the charges of the lepton doublet, as extracted from Eq.\  (\ref{eq1}), to be 
\begin{gather}
\begin{split}
\left( 
\begin{array}{c}
v_e \\ e
\end{array}
\right)_L : ~& g_L \sin\theta_w 
\left(
\begin{array}{c}
1/2 \\ -1/2
\end{array}
\right)
+ \frac{g_4 \cos\theta_w \sin\phi}{\sqrt{24}}(-3)
\\
&\qquad = \left(
\begin{array}{c}
e/2 \\ -e/2
\end{array}
\right)
+ \sqrt{\frac{2}{3}} \sqrt{\frac{1}{24}} (-3e) = 
\left(
\begin{array}{c}
0 \\ -e
\end{array}
\right)
\end{split}
\label{eq3}
\\
\begin{split}
\left( 
\begin{array}{c}
v_e \\ e
\end{array}
\right)_R: ~&g_R \cos\theta_w \cos\phi 
\left(
\begin{array}{c}
1/2 \\ -1/2
\end{array}
\right)
+ \frac{g_4 \cos\theta_w \sin\phi}{\sqrt{24}}(-3) = 
\left(
\begin{array}{c}
0 \\ -e
\end{array}
\right)
\end{split}
\label{eq4}
\end{gather}
and those of the quark doublets to be
\begin{gather}
\begin{split}
\left( 
\begin{array}{c}
v \\ d
\end{array}
\right)_L : ~&g_L \sin\theta_w 
\left(
\begin{array}{c}
1/2 \\ -1/2
\end{array}
\right)
+ \frac{g_4 \cos\theta_w \sin\phi}{\sqrt{24}}(+1)
\\
&\qquad = 
\left(
\begin{array}{c}
e/2 \\ -e/2
\end{array}
\right)
+ (e/6) = 
\left(
\begin{array}{c}
2e/3 \\ -e/3
\end{array}
\right)
\end{split}
\label{eq5}
\\
\left( 
\begin{array}{c}
v \\ d
\end{array}
\right)_R : ~g_R \cos\theta_w \cos\phi 
\left(
\begin{array}{c}
1/2 \\ -1/2
\end{array}
\right)
+ \frac{g_4 \cos\theta_w \sin\phi}{\sqrt{24}}(+1)
= 
\left(
\begin{array}{c}
2e/3 \\ -e/3
\end{array}
\right)~.
\label{eq6}
\end{gather}
These conventional charge assignments, as generated by relations (\ref{eq2}), imply that
\begin{equation}
e^2 \left[ \frac{1}{g_L^2} + \frac{1}{g_R^2} + \frac{2}{3g_4^2} \right] = 1
\label{eq7}
\end{equation}
for fractionally charged quarks, a result quoted in Eq.\ (17) of ref.\ \cite{2}.
The usual $U(1)$ coupling constant $g^\prime$ for electroweak symmetry is then given by
\begin{equation}
\frac{1}{g^{\prime 2}} = \frac{1}{g_R^2} + \frac{2}{3g_4^2}~.
\label{eq8}
\end{equation}

At present we are considering the hierarchy
\begin{equation}
\begin{array}{cc}
G & M^\prime \\ \downarrow & \\
SU(2)_L \times SU(2)_R \times SU(4) & M\\
\downarrow & \\
SU(2)_L \times U(1) \times SU(3)_c & M_z 
\end{array}
\label{eq9}
\end{equation}
We assume at the intermediate mass scale $M$ the $SU(2)$ coupling constants are smaller than that of $SU(4)$; {\it i.e.,} that $g_L (M) = g_R (M) < g_4 (M)$, consistent with evolution of a symmetrically embedded Pati-Salam sub-group $\left[ g_L (M^\prime) = g_R (M^\prime) = g_4 (M^\prime)\right]$ within a GUT $G$ at mass scale $M'$, such as $SO(10)$.  We also note to one loop order that the fermionic content to the evolution of the $SU(2)$ coupling constants $g_L(\mu)$ and $g_R(\mu)$ per generation is the same as the fermionic content of $g_4(\mu)$.  Four chiral doublets of $SU(2)_{L\rm{~or~}R}$ contributes equivalently to two non-chiral fundamental respresentations of $SU(4)$.  The same situation occurs in the calculation of Georgi, Quinn and Weinberg \cite{3}, where two fundamental representations of $SU(3)_c$ contributions equivalently to the $\beta$-function as four chiral $SU(2)_L$ doublets.  
\footnote{Following ref.\ \protect\cite{3}, we ignore here the small Higgs contribution, which will not be the same for $SU(2)_{L,R}$ and $SU(4)$.  The present one-loop calculation is not intended to be of sufficient precision for this small and symmetry-breaking mechanism-dependent Higgs contribution to affect it appreciably.}

In the hierarchy (\ref{eq9}), the $SU(3)_c$ coupling constant devolves from the $SU(4)$ coupling constant $g_4 (M)$.  Thus we can write
\begin{equation}
\frac{1}{\alpha_s(M_z)} = \frac{4\pi}{g_3^2 (M_z)} = 4\pi \left[ \frac{1}{g_4^2 (M)} - 2 \beta_3 \log \left( \frac{M}{M_z}\right) \right]~.
\label{eq10}
\end{equation}
where $\beta_3 = \left( 33 - 2n_f \right) / 48\pi^2$.  Similarly, one finds from Eq.\ (\ref{eq7}) that
\begin{equation}
\frac{1}{\alpha(M_z)}  =  4\pi \left[ \frac{1}{g_L^2(M)} + \frac{1}{g_R^2 (M)} - 2 \left( \beta_2 + \beta_1 \right) \log \frac{M}{M_z}
 +   \frac{2}{3}\left[ \frac{1}{g_4^2 (M)} - 2\beta_1 \log \frac{M}{M_z}\right] \right]
\label{eq11}
\end{equation}  
where $\beta_2 = (22 - 2n_f) / 48 \pi^2$, $\beta_1 = -2n_f/48\pi^2$.
Since $g_L (M) = g_R (M) \equiv g(M)$, this equation may be written as
\begin{equation}
\frac{1}{\alpha(M_z)} = 4\pi \left[ \frac{2}{g^2 (M)} + \frac{2}{3g_4^2 (M)} - \left( \frac{10}{3} \beta_1 + 2\beta_2 \right) \log \frac{M}{M_z}\right]~.
\label{eq12}
\end{equation}
The relationships (\ref{eq2}) can be massaged into $M$-independent relationships as well [$g_{15} (M)_z$) and $g_3 (M_z)$ devolve from $g_4(M)$]:
\begin{equation}
\begin{split}
 \frac{4\pi}{g_L^2 (M_z)} - \left( \frac{4\pi}{g_R^2 (M_z)} 
 + \frac{2}{3} \frac{4\pi}{g_{15}^2 (M_z)} \right)+ \frac{2}{3} \frac{4\pi}{g_3^2 (M_z)} 
& =  \frac{2 \sin^2 \theta_w (M_z) - 1}{\alpha(M_z)} + \frac{2}{3\alpha_s (M_z)}
\\
& =  4\pi \left[ -\frac{4}{3} \beta_3 \log \frac{M}{M_z} + \frac{10}{3} \beta_1 \log \frac{M}{M_z} - 2\beta_2 \log \frac{M}{M_z} \right] .
\end{split}
\label{eq13}
\end{equation}
Now we know low energy values for $\alpha(M_z)$, $\alpha_s(M_z)$ and $\sin^2 \theta_w (M_z)$ \cite{4}:
$\alpha(M_z) = 1/127.908 \pm 0.019$, $\alpha_s (M_z) = 0.1213 \pm 0.0018$, $\sin^2 \theta_w (M_z) = 0.2312 \pm 0.00015$.  Given these (central) values, we can solve Eq.\ (\ref{eq13}) for $\log{ \left(M/M_z\right)}$;  note that $n_f$ cancels explicitly from the final righthand side of 
Eq.\ (\ref{eq13}).  We can then use $\alpha_s (M_z)$ and $\log (M/M_z)$ in Eq.\ (\ref{eq10}) to find $g_4 (M)$ for $n_f = 6$ flavours.  Finally, we can use 
Eq.\ (\ref{eq12}) in conjunction with $\log{ \left(M/M_z\right)}$, $g_4 (M)$ and $\alpha(M_z)$ to solve for $g(M) \left( =g_{L,R} (M) \right)$.  We then find that
\begin{equation}
\begin{split}
\log{ \left(M/M_z\right)}=27.10~\Longrightarrow M  &=  5.90 \times 10^{11} M_z \cong 5.38 \times 10^{13}\, {\rm GeV}
\\
1/g_4^2 (M)  &=  3.059
\\
1/g_{L,R}^2 (M)  &=  3.498  ~.
\end{split}
\label{eq14}
\end{equation}
In other words, the unification values given in Eq.\ (\ref{eq14}) for the intermediate Pati-Salam group are able to reproduce present low-energy $(\mu = M_z)$ phenomenological values for $\alpha\left(M_z\right)$, $\alpha_s\left(M_z\right)$, and $\sin^2 \theta_w\left(M_z\right)$.

If the Pati-Salam group devolves from a group $G$, as in the hierarchy (\ref{eq9}), one finds from the relations
\begin{equation}
\frac{1}{g_{GUT}^2 (M^\prime)}  =  \frac{1}{g_4^2 (M)} + 2\beta_4 \log \left( \frac{M^\prime}{M}\right)
 =  \frac{1}{g_{L,R}^2 (M)} + 2\beta_2 \log \left( \frac{M^\prime}{M}\right)
\label{eq15}
\end{equation}
that $\log (M^\prime / M) = 4.72$, $g_{GUT}^{-2} (M^\prime) = 3.697$, in which case the ultimate unification mass scale is $M^\prime = 6.05 \times 10^{16}\, {\rm GeV}$.

\begin{figure}[hbt]
\centering
\includegraphics[scale=0.5]{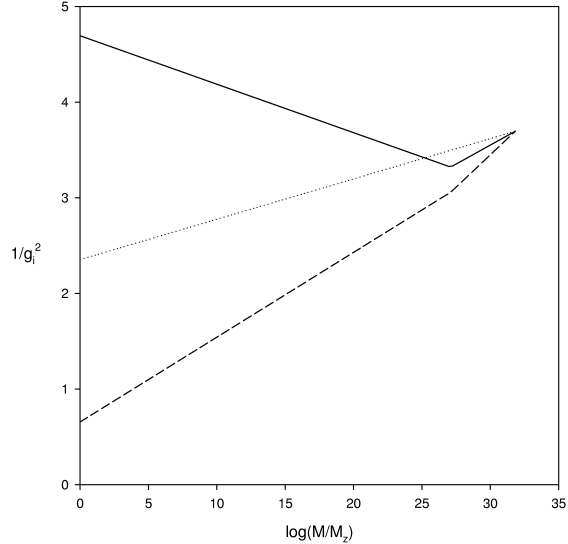}
\caption{One loop evolution of the couplings from the scale $M_z$ to the unification scale $M'$ where the three couplants merge.  The solid curve represents $g_1$, the dotted curve represents $g_2$, and the dashed curve represents $g_3$.  
}
\label{fig1}
\end{figure}

\begin{figure}[hbt]
\centering
\includegraphics[scale=0.5]{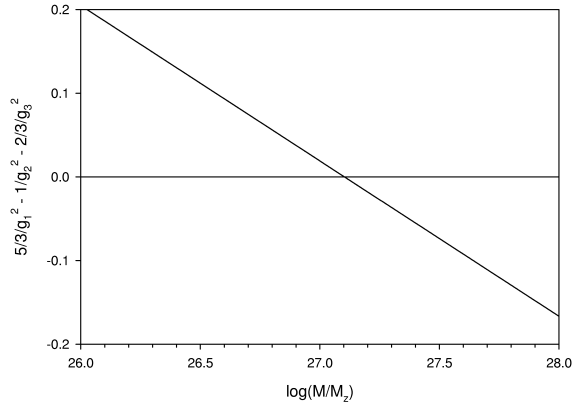}
\caption{One-loop evolution of the combination $5/3 g_1^{-2}-g_2^{-2}-2/3g_3^{-2}$.  The Pati-Salam scale $M$ is identified as the point where the curve passes through zero, and is in numerical agreement with the value $\log\left(M/M_z\right)=27.10$ in Eq.~(\ref{eq14}).
}
\label{fig_oneloop}
\end{figure}

The one-loop evaluation of coupling constants is depicted in Fig.\ \ref{fig1}.  Note that the couplants all converge at the GUT scale $M'$.  Before reaching the scale $M$ for $G_{PS}=SU(2)_L\times SU(2)_R\times SU(4)_C$ symmetry, the $U(1)$ inverse couplant
$g_1^{-2}$ becomes somewhat smaller than the $SU(2)$ inverse couplant $g_2^{-2}$.  This is a reflection of the 
following constraint satisfied by the couplings at the intermediate scale $M$ \cite{2}:
\begin{equation}
g_1^{-2}(M)=\frac{3}{5}\left[ g_2^{-2}(M) +\frac{2}{3} g_3^{-2}(M)\right]~.
\label{neweq2}
\end{equation}
Indeed, the Pati-Salam scale $M$ can be identified  as the scale at which $g_1^{-2}(\mu)$ evolves from $g_1^{-2}\left(M_z\right)$ to the right-hand side of Eq.~(\ref{neweq2}), as determined by evolution from $g_2^{-2}\left(M_z\right)$ and $g_3^{-2}\left(M_z\right)$.  This is illustrated in Fig.~\ref{fig_oneloop} in which $\log{\left(M/M_z\right)}$ is determined to be the $x$-intercept of the curve.

Similarly we note that $g_1^{-2}(\mu)$ must remain above $g_3^{-2}(\mu)$ as in Fig.~\ref{fig1}.  In the range $M<\mu<M'$  we have
$g_1^{-2}(\mu)>g_3^{-2}(\mu)$ since the evolution of $g_1^{-2}$ contains a component proportional to $g_2^{-2}(\mu)$ whereas 
$g_3^{-2}(\mu)$ involves entirely like $g_4^{-2}(\mu)$ in the range $M<\mu<M'$ [$SU(3)_c$ is embedded entirely within $SU(4)_c$]. 
Consequently, if $g_1^{-2}(M)<g_3^{-2}(M)$ coupling constant unification at a scale higher than $M$ is necessarily impossible.

\begin{figure}[hbt]
\centering
\includegraphics[scale=0.5]{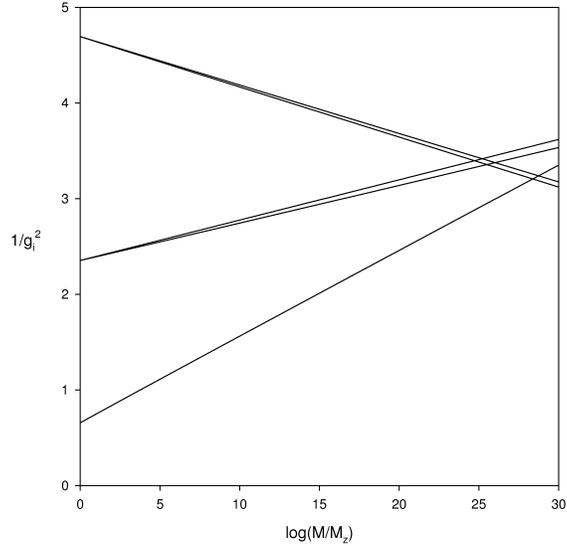}
\caption{Comparison of one-loop and two-loop evolution of the gauge couplings for $M_H=220\,{\rm GeV}$.  
The upper curves represent $g_1$, the middle curves represent $g_2$, and the bottom curves (which overlap within the resolution of the plot) represent $g_3$.  
}
\label{fig_comp}
\end{figure}

\begin{figure}[hbt]
\centering
\includegraphics[scale=0.5]{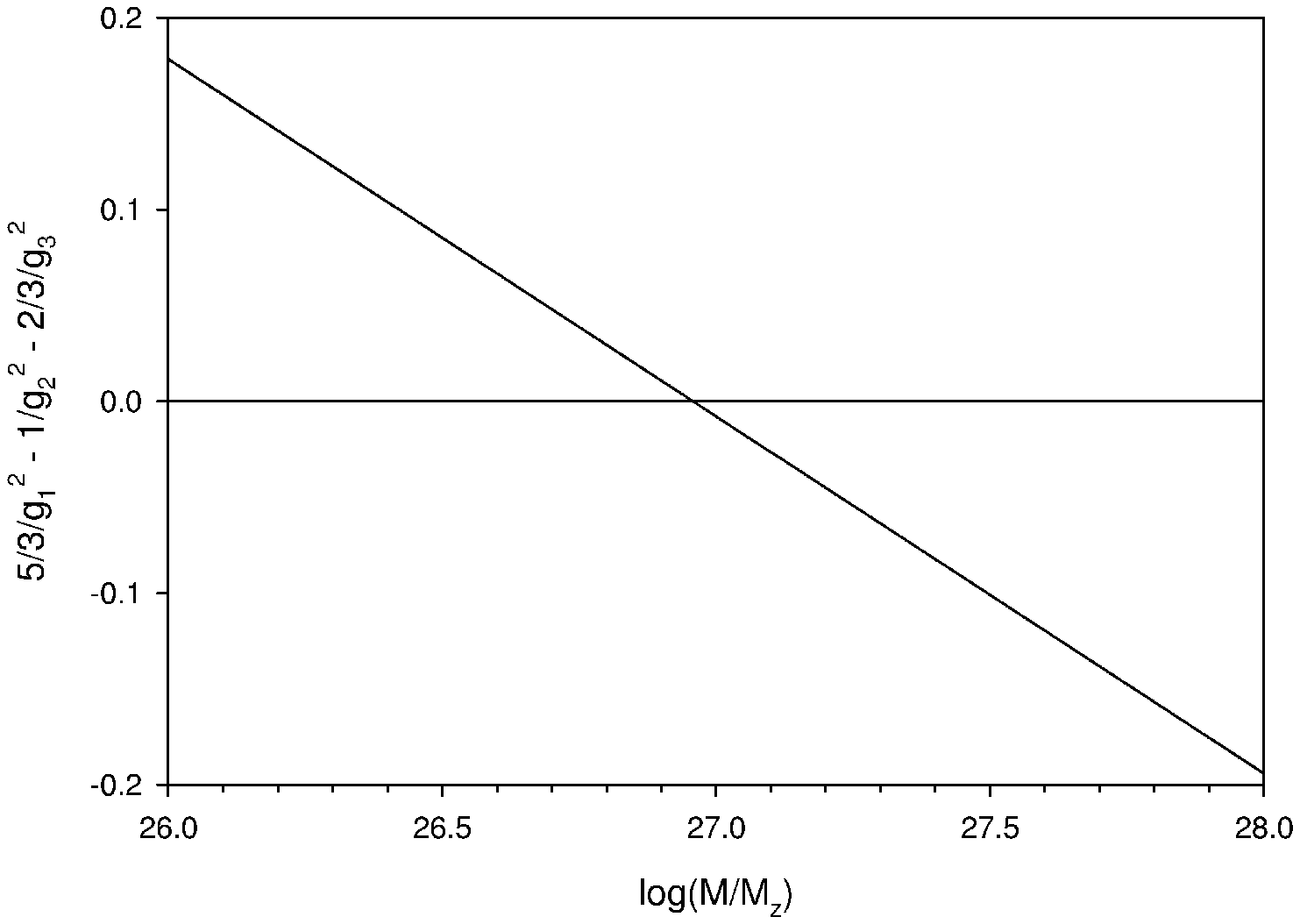}
\caption{Two-loop evolution of the combination $5/3 g_1^{-2}-g_2^{-2}-2/3g_3^{-2}$ for $M_H=220\,{\rm GeV}$.  The Pati-Salam scale $M$ is identified as the point where the curve passes through zero.).
}
\label{fig_twoloop}
\end{figure}

Because of the narrow window in which $g_3^{-2}<g_1^{-2}<g_2^{-2}$, one has to question the stability of the one-loop result.  The significant known interaction contributions to the one-loop hierarchy are given by the corresponding two-loop $\beta$-functions \cite{ford}
{\allowdisplaybreaks
\begin{gather}
\mu\frac{dg_3}{d\mu}=\frac{1}{16\pi^2} \left(-7g_3^3\right)+\frac{1}{\left(16\pi^2\right)^2}
\left[-26g_3^5-2h^2g_3^3+\frac{11}{10}g_1^2g_3^3+\frac{9}{2}g_2^2g_3^3\right]
\\
\mu\frac{dg_2}{d\mu}=\frac{1}{16\pi^2} \left(-\frac{19}{6}g_2^3\right)+\frac{1}{\left(16\pi^2\right)^2}
\left[ \frac{35}{6}g_2^5+12g_3^2g_2^3+\frac{9}{10}g_1^2g_2^3-\frac{3}{2}h^2g_2^3   \right]
\\
\mu\frac{dg_1}{d\mu}=\frac{1}{16\pi^2} \left(\frac{41}{10}g_1^3\right)+\frac{1}{\left(16\pi^2\right)^2}
\left[\frac{199}{50}g_1^5+\frac{27}{10}g_2^2g_1^3+\frac{44}{5}g_3^2g_1^3-\frac{17}{10}g_1^3h^2  \right]
\\
\mu\frac{dh}{d\mu}=\frac{1}{16\pi^2} \left( \frac{9}{2}h^3-8g_3^2h-\frac{9}{4}g_2^2h-\frac{17}{10}g_1^2h \right)
+\frac{1}{\left(16\pi^2\right)^2}
\left[-12h^5-12\lambda h^3+\lambda^2h+36 g_3^2 h^3-108 g_3^4 h \right]
\\
\begin{split}
\mu\frac{d\lambda}{d\mu}=&\frac{1}{16\pi^2} \left(24\lambda^2+12\lambda h^2-6h^4-9\lambda g_2^2-\frac{9}{5}\lambda g_1^2+\frac{27}{200}g_1^4+\frac{9}{20}g_1^2g_2^2+\frac{9}{8}g_2^4\right)
\\
&+\frac{1}{\left(16\pi^2\right)^2}
\left[-312\lambda^3-144 \lambda^2 h^2-3\lambda h^4+30 h^6+80 \lambda g_3^2h^2-32h^4g_3^2 \right]~,
\end{split}
\end{gather}
}
where $h$ is the $t$-quark Yukawa coupling $h\left(m_t\right)\cong h\left(M_z\right)\cong 1.00$ and where $\lambda$ is the quartic scalar coupling $\lambda\left(M_H\right)\cong \lambda\left(M_z\right)\cong M_H^2/2v^2$ for 
conventional symmetry breaking.\footnote{Note that our $\lambda$ is 6 times that in Ref.~\cite{ford} because of a different numerical coefficient in the $\phi^4$ term in the Standard Model potential.  In our notation the $\phi^4$ coefficient is $\lambda/4$ as opposed to $\lambda/24$ in \cite{ford}.}
The coupling $\lambda$, whose numerical value at $M_z$ depends on the Higgs mass, feeds into the evolution of the gauge couplings 
$\{g_1,g_2,g_3\}$---it does not contribute directly to $g_2$ and $g_1$ $\beta$-functions until three-loop order.
However, $\lambda$ does enter  the leading evolution of $h(\mu)$ which in turn enters the two-loop $\beta$-functions of
$\{g_1,g_2,g_3\}$ and could conceivably upset the balance by which $g_3^{-2}(M)<g_1^{-2}(M)<g_2^{-2}(M)$. 
Within the range $150\,{\rm GeV}<M_H<300\,{\rm GeV}$ we find in going from one-loop to two-loop order virtually identical evolution curves (see Figure \ref{fig_comp}).  Similarly, the two-loop solution of the numerical constraint (\ref{neweq2}) leading to the $G_{PS}$ intermediate scale results in only a slight reduction to $\log\left(M/M_z\right)=27.0$ (see Figure \ref{fig_twoloop}).

Thus empirical bounds on the intermediate mass scale $M$ are not appreciably altered in going from one-loop to two-loop order, and  a single intermediate symmetry $G_{PS}$ in the group hierarchy (\ref{eq9}) retains its viability.  
We have therefore found  that an intermediate Pati-Salam level of symmetry, as in the hierarchy (\ref{eq9}), can accommodate known ``low-energy'' values  $\alpha\left(M_z\right)$, $\alpha_s\left(M_z\right)$ and $\sin^2 \theta_w\left(M_z\right)$ even if the evolution of couplings is considered to two-loop order. Moreover, the intermediate mass scale $M$ does not change appreciably in moving from one- to two-loop order [$\log{\left(M/M_Z\right)}\approx 27$]. Since the intermediate mass scale $M$ is so very large, however, there remain severe hierarchy problems in this scenario.  For example, one might have to appeal to a radiative breaking of electroweak symmetry \cite{5}, with a concomitant exact symmetry preventing $\phi^2$ terms in the Higgs potential, in order to avoid contamination of the Higgs mass from scales $M$ and $M^\prime$. We note in conclusion, though, that gauge coupling constant unification is seen to occur in a sensible hierarchy {\it without} supersymmetry, as is generally assumed to be required, a scenario also not entirely immune from hierarchy problems.

We are grateful for support from the Natural Science and Research Council of Canada.


\end{document}